\documentclass[12pt]{article}

\usepackage{graphics,epsf}

\usepackage{amsfonts}
\usepackage[footnotesize,bf]{caption}

\newcommand{\nit}{\noindent}

\newcommand{\np}{\newpage}
\newcommand{\dsp}{\displaystyle}
\newcommand{\vs}[1]{\vspace{#1 ex}}
\newcommand{\hs}[1]{\hspace{#1 em}}
\newcommand{\bflr}{\begin{flushright}}
\newcommand{\eflr}{\end{flushright}}
\newcommand{\bc}{\begin{center}}
\newcommand{\ec}{\end{center}}
\newcommand{\ben}{\begin{enumerate}}
\newcommand{\een}{\end{enumerate}}

\newcommand{\be}{\begin{equation}}
\newcommand{\ee}{\end{equation}}
\newcommand{\ba}{\begin{array}}
\newcommand{\ea}{\end{array}}
\newcommand{\ct}{\cite}
\newcommand{\bit}{\bibitem}
\newcommand{\dd}[2]{\frac{\partial{#1}}{\partial{#2}}}
\newcommand{\ag}{\alpha}
\newcommand{\bg}{\beta}
\newcommand{\gam}{\gamma}
\newcommand{\del}{\delta}
\newcommand{\eps}{\epsilon}
\newcommand{\ve}{\varepsilon}

\newcommand{\thg}{\theta}
\newcommand{\kg}{\kappa}

\newcommand{\rg}{\rho}

\newcommand{\og}{\omega}
\newcommand{\Gam}{\Gamma}
\newcommand{\Del}{\Delta}
\newcommand{\Fg}{\Phi}

\newcommand{\Lb}{\Lambda}

\newcommand{\bfkg}{\mbox{{\boldmath $\kg$}}}

\newcommand{\bfa}{{\bf a}}

\newcommand{\bfj}{{\bf j}}
\newcommand{\bfk}{{\bf k}}

\newcommand{\bfq}{{\bf q}}

\newcommand{\bfu}{{\bf u}}

\newcommand{\bfx}{{\bf x}}
\newcommand{\bfy}{{\bf y}}

\newcommand{\bfG}{{\bf G}}

\newcommand{\bfI}{{\bf I}}

\newcommand{\bfP}{{\bf P}}
\newcommand{\bfR}{{\bf R}}

\newcommand{\lh}{\left(}
\newcommand{\rh}{\right)}
\newcommand{\ld}{\left.}

\newcommand{\nb}{\nabla}
\newcommand{\bfnb}{\mbox{\boldmath{$\nb$}}}

\newcommand{\lrder}{\stackrel{\leftrightarrow}{\nabla}}
\newcommand{\bflrder}{\stackrel{\leftrightarrow}{\bfnb}}

\newcommand{\cN}{{\cal N}}

\newcommand{\der}{\partial}

\begin{document}

\pagestyle{empty}
\begin{flushright}
NIKHEF/2007-026
\end{flushright}

\bc
{\Large {\bf The fate of conformal symmetry}} \\
\vs{2}

{\Large {\bf  in the non-linear Schr\"{o}dinger theory}} \\
\vs{5}

{\large M.O.\ de Kok} 
\vs{2}

Inst.\ Lorentz, University of Leiden 
\vs{1}

P.O.\ Box 9506 
\vs{1}

2300 RA Leiden NL
\vs{1}

e-mail: mdekok@lorentz.leidenuniv.nl
\vs{3}

{\large J.W.\ van Holten} 
\vs{2}

Nikhef
\vs{1}

P.O.\ Box 41882
\vs{1}

1009 DB Amsterdam NL 
\vs{1}

e-mail: v.holten@nikhef.nl
\vs{3}

\today
\ec
\vs{5}

{\footnotesize
\nit
{\bf Abstract}\\
The free Schr\"{o}dinger theory in $d$ space dimensions is a 
non-relativistic conformal field theory. The interacting 
non-linear theory preserves this symmetry in specific numbers
of dimensions at the classical (tree) level.  This holds in
particular for the $|\Fg|^4$-theory in $d = 2$. We compute the
full quantum corrections to the 1PI 4-point function in $d = 2 - \eps$  
dimensions and find a non-trivial $\bg$-function completely given 
by the 1-loop result. We exhibit an explicit Ward-identity showing 
that scale-invariance is broken in the limit $d = 2$ by an anomalous 
contribution proportional to the $\bg$-function. }

\np
\pagestyle{plain}
\pagenumbering{arabic}

\nit
{\bf 1.\ Conformal symmetry of the free Schr\"{od}inger theory} 
\vs{1}

\nit
Conformal symmetry, in exact or broken form, plays an important role 
in field theory. As it governs the scale dependence of a theory, it 
implicitly determines the regime of applicability of the theory. 
String theory, as an important candidate for a unified theory of 
quantum gravity, has an exact conformal invariance when formulated 
as a 2-$D$ field theory on the world sheet. On the other hand, 
effective field theories in four space-time dimensions, like QED or 
the standard model, possess an approximate scale invariance, broken 
explicitly by mass terms and/or by quantum effects. 
Indeed, very few theories are known to be exactly scale invariant,
except in 2 dimensions \ct{BPZ}. In 4 dimensions $N = 4$ Yang-Mills
theory is known to have vanishing $\bg$-function to all orders 
\ct{mandelstam}, but in general gauge theories which are classically 
scale invariant, such as QCD with massless quarks, have this symmetry 
broken at the quantum level. 

In this paper we consider conformal symmetry in the context of 
non-relativistic field theory. Such symmetries were first identified 
in \ct{hagen, niederer}. The symmetries and their realization in 
classical and quantum field theory have been studied by several 
authors \ct{jackiw-pi}-\ct{nishida-son}. We consider in particular 
theories describing Bose gases in $d$ space dimensions. 
Conformal symmetry is an exact symmetry of the free theory, but it 
can be implemented in certain interacting models as well. The 
quantum field theoretical aspects of Bose gases in general have 
been studied widely in the literature; reviews and references can 
be found e.g.\ in \ct{abrikosov, stoof}.
 
In $d$ space dimensions the free theory is defined by the action
\[
S_0 = \int dt \int d^d x \lh \frac{i}{2}\, \Psi^* 
 \stackrel{\leftrightarrow}{\der}_t \Psi 
 - \frac{1}{2m}\, \bfnb \Psi^* \cdot \bfnb \Psi \rh.
\]
Without loss of generality the theory can be simplified by rescaling 
the time variable
\[
\tau = \frac{t}{m},
\]
which is equivalent to choosing units in which $m = 1$. The action
then reads
\be 
S_0 = \int d\tau \int d^d x \lh \frac{i}{2}\, \Psi^* 
 \stackrel{\leftrightarrow}{\der}_{\tau} \Psi 
 - \frac{1}{2}\, \bfnb \Psi^* \cdot \bfnb \Psi \rh,
\label{1.1}
\ee 
which is stationary when $\Psi$ satisfies the linear Schr\"{o}dinger
equation for free particles with unit mass:
\be
i \der_{\tau} \Psi = - \frac{1}{2}\, \Del \Psi.
\label{1.2}
\ee 
This equation and the action (\ref{1.1}) are invariant under the 
Schr\"{o}dinger group of space-time transformations 
\ct{niederer,henkeluu}, which includes time and space translations, 
spatial rotations, Galilei boosts, dilatations and special conformal 
transformations. The explicit form of these transformations is presented 
in the appendix. The associated constants of motion are: \\ 
a.\ the hamiltonian 
\be
H_0 = \frac{1}{2}\, \int d^d x\, \bfnb \Psi^* \cdot \bfnb \Psi;
\label{v.1}
\ee 
b.\  the momentum 
\be
\bfP = \frac{i}{2}\, \int d^dx\, \Psi^* \bflrder \Psi;
\label{v.2}
\ee
c.\ the angular momentum
\be
M_{ij} = \frac{i}{2}\, \int d^d x \lh x_i (\Psi^* \lrder_j \Psi)
 - x_j (\Psi^* \lrder_i \Psi) \rh;
\label{v.3}
\ee
d.\ the galilean boost operator  
\be 
\bfG = \tau \bfP + \int d^dx\, \bfx\, \Psi^* \Psi;
\label{v.4}
\ee
e.\ the scaling operator
\be
D_0 = 2 \tau H_0 + \frac{i}{2}\, 
 \int d^d x\, \bfx \cdot (\Psi^* \bflrder \Psi);
\label{v.5}
\ee
f.\ the special conformal charge
\be 
K_0 = \tau^2 H_0 - \tau D_0 - \frac{1}{2}\, \int d^d x\, \bfx^2\, \Psi^* \Psi;
\label{v.6}
\ee
g.\ the particle number 
\be 
N = \int d^d x\, \Psi^* \Psi.
\label{v.7}
\ee  
By the fundamental Poisson bracket 
\be 
\left\{ \Psi(\bfx,\tau), \Psi^*(\bfy,\tau) \right\} = -i \del^d(\bfx - \bfy),
\label{1.3}
\ee
the constants of motion (a-g) generate the infinitesimal transformations 
of the fields $\Psi$ and $\Psi^*$ which leave the action (\ref{1.1}) 
invariant. As a result they realize the Schr\"{o}dinger algebra
extended with a central charge $N$: 
\be
\ba{l}
\left\{ M_{ij}, M_{kl} \right\} = \del_{il} M_{jk} - \del_{ik} M_{jl}
 - \del_{jl} M_{ik} + \del_{jk} M_{il}, \\
 \\ 
\left\{ H_0, D_0 \right\} = 2 H_0, \hs{2.9} \left\{ H_0, \bfG \right\} = \bfP, \\
 \\ 
\left\{ \bfP, D_0 \right\} = \bfP, \hs{3.5}
\left\{ P_i, M_{jk} \right\} = \del_{ij}\, P_k - \del_{ik}, P_j, \\
 \\
\left\{ H_0, K_0 \right\} = - D_0, \hs{2.5} 
\left\{ P_i, G_j \right\} = \del_{ij}\, N,  
\\
  \\
\left\{\bfG, D_0 \right\} = - \bfG, \hs{2.5} 
\left\{ G_i, M_{jk} \right\} = \del_{ij} G_k - \del_{ik} G_j, \\
 \\
\left\{ K_0, D_0 \right\} = -2K_0, \hs{2} \left\{ K_0, \bfP \right\} = \bfG, 
\ea
\label{1.4}
\ee
all other brackets vanishing. Notice, that the central charge is 
the particle number $N$, which generates a phase transformation of 
the complex field $\Psi$. 
\vs{1}

\nit
Equivalently, the theory can be defined by the hamiltonian (\ref{v.1}) 
generating the time evolution of phase-space functions 
$F[\Psi,\Psi^*;\tau]$: 
\be 
\frac{dF}{d\tau} = \ld \dd{F}{\tau}\right|_{\Psi,\Psi^*} 
 + \left\{ F, H_0 \right\},
\label{1.5}
\ee
where the partial time derivative refers to time dependence other
than through $(\Psi, \Psi^*)$. For the constants of motion this 
implies directly that
\be
\left\{ \bfP, H_0 \right\} = \left\{ M_{ij}, H_0 \right\} 
 = \left\{ N, H_0 \right\} = 0,
\label{1.6}
\ee 
whilst the explicit time dependence in the definition of $\bfG$, $D_0$
and $K_0$ implies  
\be
\left\{ \bfG, H_0 \right\} = - \bfP, \hs{2}
\left\{ D_0, H_0 \right\} = - 2H_0, \hs{2}
\left\{ K_0, H_0 \right\} = D_0,
\label{1.7}
\ee
in agreement with eqs.\ (\ref{1.4}).

The conformal symmetry has direct relevance for the physical content
of the theory. In particular, consider the quantities \ct{ghosh}
\be
\ba{l}
\bfI \equiv \dsp{ \int d^d x\, \bfx\, \Psi^* \Psi 
 = \bfG - \tau \bfP, }\\
 \\
I_1 \equiv \dsp{ \frac{1}{2}\, \int d^d x\, \bfx^2\, \Psi^* \Psi 
 = \tau^2 H_0 - \tau D_0 - K_0, }\\
 \\
I_2 \equiv \dsp{ \frac{i}{2}\, \int d^d x\, \bfx \cdot \Psi^* \bflrder \Psi 
 = D_0 - 2\tau H_0.}
\ea
\label{1.8}
\ee
As $(H_0, \bfP, \bfG, D_0, K_0)$ are constants of motion, it follows 
that 
\be 
\frac{d\bfI}{d\tau} = - \bfP, \hs{2}
\frac{dI_1}{d\tau} = - I_2, \hs{2} \frac{dI_2}{d\tau} = - 2H_0.
\label{1.9}
\ee
Therefore, if there would be any {\em static} (i.e.\ time-independent)
solutions of the theory, such that $(\bfI, I_1, I_2)$ are constant 
themselves, they would necessarily have zero energy and momentum. 
Of course, such solutions don't exist in the present case: all the 
solutions of the free Schr\"{o}dinger theory represent scattering 
states, superpositions of $\del$-function normalizable plane waves
with a continuous energy spectrum. 
\vs{2}

\nit
{\bf 2.\ Conformal symmetry in the non-linear Schr\"{o}dinger model} 
\vs{1}

\nit
In general the conformal transformations which are part of the 
Schr\"{o}dinger group and a symmetry of the free theory, are broken 
by interactions. In particular, $U(1)$-invariant polynomial interactions 
of the form
\be
S_{int} = - \frac{g^2}{n}\, \int d\tau \int d^dx\, \lh \Psi^* \Psi \rh^n,
\label{2.1}
\ee
where $g$ is a coupling constant, break the scale invariance unless 
$n$ is such that the dimension of $|\Psi|^{2n}$ matches\footnote{The
Weyl weight of the field $\Psi$ being $d/2$; see eq.\ (\ref{a2.9})} 
that of the space-time integration measure:
\be
nd = d + 2,
\label{2.2}
\ee
implying that $g$ is dimensionless. In the case $d = 1$ this is satified for 
$n= 3$, in the case $d = 2$ for $n = 2$; also, for $d \rightarrow \infty$ 
we get $n \rightarrow 1$. In all other cases $n$ must take on non-integer 
values. With the interactions included, the hamiltonian becomes
\be
H_n = \int d^d x \lh \frac{1}{2}\, \bfnb \Psi^* \cdot \bfnb \Psi
 + \frac{g^2}{n}\, \lh \Psi^* \Psi \rh^n \rh.
\label{2.3}
\ee 
Systems described by such a hamiltonian appear frequently in the context 
of condensed matter systems \ct{courteille}, where the corresponding 
non-linear Schr\"{o}dinger equation
\be 
i \der_{\tau} \Psi = - \frac{1}{2}\, \Del \Psi + g^2 |\Psi|^{n-1} \Psi,
\label{2.3.1}
\ee
is also known as the Gross-Pitaevskii equation \ct{pitaevskii, gross}. 

The non-linear theory defined by eqs.\ (\ref{2.1}), (\ref{2.3}) is
still invariant under time- and space-translations, spatial rotations 
and $U(1)$ phase transformations. As a result the generators 
$(\bfP, M_{ij}, \bfG, N)$ as in (\ref{v.2})-(\ref{v.7}) also represent 
constants of motion in the interacting theory. The generators of scale 
and special conformal transformations are replaced by
\be
\ba{lll}
D_n & = & 2 \tau H_n + I_2, \\
 & & \\
K_n & = & \tau^2 H_n - \tau D_n - I_1,
\ea
\label{2.3.2}
\ee
with $(I_1, I_2)$ defined by the explicit first expressions in eqs.\ 
(\ref{1.8}). They indeed generate the infinitesimal transformations 
\be
\ba{lll}
\left\{ \Psi, D_n \right\} & = & \dsp{ \lh 2 \tau \der_{\tau} 
 + \bfx \cdot \bfnb + \frac{d}{2} \rh \Psi, }\\ 
 & & \\
\left\{ \Psi, K_n \right\} & = & \dsp{ \lh - \tau^2 \der_{\tau} 
 - \tau \lh \frac{d}{2} + \bfx \cdot \bfnb \rh 
 + \frac{i}{2}\, \bfx^2 \rh \Psi. }
\ea
\label{2.5}
\ee
However, $D_n$ and $K_n$ are constants of motion only if $n$ and $d$
are related by (\ref{2.2}). To show this, we note the results  
\be 
\ba{l}
\left\{ \bfI, H_n \right\} = - \bfP, \hs{2}
\left\{ I_1, H_n \right\} = - I_2, \\
 \\
\dsp{ \left\{ I_2, H_n \right\} = - 2 H_n 
 + \frac{(2 + d - nd)}{n}\, \int d^d x\, g^2 \lh \Psi^* \Psi \rh^n.}
\ea
\label{2.6}
\ee 
The first of these equations implies the conservation of $\bfG$ for 
all $n$ and $d$; the other two equations imply 
\be 
\frac{dK_n}{d\tau} = - \tau \frac{dD_n}{d\tau}, \hs{2}
\frac{dD_n}{d\tau} = \frac{(2 + d - nd)}{n}\, 
 \int d^dx\, g^2\, \lh \Psi^* \Psi \rh^n. 
\label{2.7}
\ee 
Eq.\ (\ref{2.2}) is indeed the necessary and sufficient condition
for the right-hand side of these equations to vanish. In these cases 
we obtain an interacting classical field theory invariant under the 
full Schr\"{o}dinger group of tranformations. For these theories one 
can derive equations similar to (\ref{1.9}):
\be
\frac{d\bfI}{d\tau} = - \bfP, \hs{2} \frac{dI_1}{d\tau} = -I_2, \hs{2}
\frac{dI_2}{d\tau} = - 2 H_n,
\label{2.4}
\ee
and conclude that again any static solutions of the interacting theories 
have zero energy and momentum. Now it is obvious, that both the free 
hamiltonian $H_0$ as well as $H_n$ in the interacting theory with 
$g^2 > 0$ are non-negative. Therefore the only zero-energy solution is 
$\Psi = 0$; it follows immediately, that the conformal models have no 
non-trivial static solutions, neither the $|\Psi|^6$-model in $d = 1$, 
nor the $|\Psi|^4$-model in $d = 2$. As in the free theory, the physical
solutions represent scattering states with a continuous energy spectrum. 
A more complete discussion is presented in appendix B. 
\vs{2}

\nit
{\bf 3.\ Quantum non-linear Schr\"{o}dinger model}
\vs{1}

\nit
In the remainder of this paper we consider the non-linear 
Schr\"{o}dinger model with $|\Psi|^4$ interactions. As we
have seen, the classical field theory is conformally invariant
in $d=2$ space dimensions. The problem we wish to address is 
whether the conformal symmetry is preserved in the corresponding
quantum field theory. 

The quantum field theory is obtained by taking $\Psi$ and $\Psi^*$ 
to be conjugate operators with the equal-time commutation relation 
\be
\left[ \Psi(\bfx, \tau), \Psi^*(\bfy, \tau) \right] = \del^d (\bfx - \bfy).
\label{3.1}
\ee
Furthermore we take the hamiltonian and other generators of the
Schr\"{o}dinger group to be the normal-ordered operator version of 
the corresponding classical ones:
\be 
H_{\mu}[\Psi,\Psi^*] = \int d^dx\, 
 \lh \frac{1}{2}\, \bfnb \Psi^* \cdot \bfnb \Psi 
 + \mu \Psi^* \Psi + \frac{g^2}{2}\, \Psi^{*\,2} \Psi^2 \rh;
\label{3.2}
\ee
The bilinear term $\mu \Psi^* \Psi$ arises from the operator-ordering
ambiguity in going from the classical to the quantum theory. As in the
classical theory there exist conserved operators $(\bfP, M_{ij}, \bfG, N)$, 
defined by the expressions (\ref{v.2})-(\ref{v.7}) with $(\Psi, \Psi^*)$ 
interpreted as field operators. It follows immediately, that we can 
subtract a constant term $\mu N$ from the energy and define another
$\mu$-independent conserved energy operator
\be
H_s = H - \mu N = \int d^d x \lh  \frac{1}{2}\, \bfnb \Psi^* \cdot \bfnb \Psi 
 + \frac{g^2}{2}\, \Psi^{*\,2} \Psi^2 \rh.
\label{3.2.1}
\ee 
Equivalently, we can redefine the fields: 
\be
\Fg(\bfx,t) = e^{-i\mu t} \Psi(\bfx,t), \hs{2} 
\Fg^*(\bfx,t) = e^{i\mu t} \Psi^*(\bfx,t),
\label{3s.1}
\ee
with the same equal-time commutation relations
\be 
\left[ \Fg(\bfx,t), \Fg^*(\bfy,t) \right] = \del^d(\bfx - \bfy).
\label{3s.2}
\ee
In terms of these fields the hamiltonian (\ref{3.2}) becomes
\be 
H[\Fg,\Fg^*] = \int d^dx\, \lh \frac{1}{2}\, \bfnb \Fg^* \cdot \bfnb \Fg 
 + \frac{g^2}{2}\, \Fg^{*\,2} \Fg^2 \rh;
\label{3s.3}
\ee
More generally, a shift in the value of $\mu$ in the hamiltonian 
(\ref{3.2}) can be compensated by a multiplicative field renormalization 
of the type (\ref{3s.1}).

Assuming such a (finite) renormalization of the fields and the $\mu$
term to have been performed, we consider the theory as defined by 
(\ref{3s.2}) and (\ref{3s.3}). In addition to the conserved operators
of the galilean transformations $(\bfP, M_{ij}, \bfG, N)$, which
take the same form in terms of the new fields $(\Fg, \Fg^*)$, we
can then also construct the conformal operators
\be
\ba{l}
D = 2 \tau H + I_2, \\ 
\\
K = \tau^2 H - \tau D - I_1, 
\ea
\label{3.2.2}
\ee 
where $I_{1,2}$ are the normal-ordered expressions (\ref{1.8}). 

The transformations of the field operators under the full Schr\"{o}dinger
algebra are obtained by computing their commutator with the generators:
\be 
\ba{l}
- i \left[ \Fg, H \right] = \der_{\tau} \Fg, \hs{10}
- i \left[ \Fg, \bfP \right] = \bfnb \Fg, \\
 \\
- i \left[ \Fg, M_{ij} \right] = \lh x_i \nb_j - x_j \nb_i \rh \Fg, \hs{3.9}
- i \left[ \Fg, \bfG \right] = \lh \tau \bfnb - i \bfx \rh \Fg, \\
 \\
- i \left[ \Fg, D \right] = \dsp{ \lh 2 \tau \der_{\tau} 
 + \bfx \cdot \bfnb + \frac{d}{2} \rh \Fg, } \hs{1.6}
- i \left[ \Fg, N \right] = -i \Fg, \\
 \\ 
- i \left[ \Fg, K \right] = \dsp{ \lh \tau^2 \der_{\tau} 
 - \tau \lh \bfx \cdot \bfnb + \frac{d}{2} \rh 
 + \frac{i}{2}\, \bfx^2 \rh \Fg. }
\ea
\label{3.3}
\ee
For any operator $F[\Fg, \Fg^*;\tau]$ the Heisenberg equation of motion 
now reads
\be
\frac{dF}{d\tau} = \ld \dd{F}{\tau} \right|_{\Fg,\Fg^*} - i \left[ F, H \right].
\label{3.4}
\ee
As in the classical theory, for the full set of Schr\"{o}dinger operators 
to be conserved: $dF/d\tau = 0$,  imposes additional constraints. Indeed, 
we reobtain the results (\ref{2.7}) as normal-ordered operator equations:
\be
\frac{dK}{d\tau} = - \tau \frac{dD}{d\tau}, \hs{2}
\frac{dD}{d\tau} = \frac{(2 - d)}{2}\, g^2\, \int d^dx\, \Fg^{*\,2} \Fg^2. 
\label{3.5}
\ee 
The relation between symmetries and constants of motion in the non-linear 
Schr\"{o}dinger theory is a specific case of Noether's general theorem. It
can be formulated locally in terms of a charge and current density 
$(\rg, \bfj)$, which for scale transformations take the form
\be 
\rg = \tau \lh \bfnb \Fg^* \cdot \bfnb \Fg + g^2 \Fg^{*\,2} \Fg^2 \rh
 + \frac{i}{2}\, \bfx \cdot \lh \Fg^* \bflrder \Fg \rh,
\label{3.6}
\ee  
and
\be 
\ba{lll}
\bfj & = & \dsp{ - \frac{d}{4}\, \bfnb \lh \Fg^* \Fg \rh - \frac{1}{2} 
 \lh \bfx \cdot \bfnb \Fg^* \bfnb \Fg + \bfnb \Fg^*\, \bfx \cdot \bfnb \Fg \rh 
 }\\
 & & \\
 & & \dsp{ +\, \frac{1}{2}\, \bfx \lh \frac{1}{2}\, \Del \lh \Fg^* \Fg \rh
  - g^2\, \Fg^{*\,2} \Fg^2 \rh }\\ 
 & & \\ 
 & & \dsp{ - \frac{i\tau}{2} \lh \bfnb \Fg^* \Del \Fg - \Del \Fg^* \bfnb \Fg 
  + g^2  \lh \Fg^{*\,2} \bflrder \Fg^2 \rh \rh. }
\ea
\label{3.7}
\ee
The operator equations of motion 
\be 
i \der_{\tau} \Fg = \left[ \Fg, H \right] = 
 - \frac{1}{2}\, \Del \Fg + g^2\, \Fg^* \Fg^2,
\label{3.8}
\ee
and its hermitean conjugate, then imply the equation of continuity
\be 
\der_{\tau} \rg + \bfnb \cdot \bfj = \frac{(2-d)}{2}\, g^2\, \Fg^{*\,2} \Fg^2.
\label{3.9}
\ee
By taking the integral over 2-dimensional space and in the absence of
boundary terms one then directly reproduces the conservation law 
(\ref{3.5}):  
\be
D = \int d^d x\, \rg \hs{1} \Rightarrow \hs{1}
\frac{d D}{d\tau} = \frac{(2-d)}{2}\, g^2 \int d^dx\, \Fg^{*\,2} \Fg^2
 = (2 - d)\, H_{int},
\label{3.10}
\ee 
where $H_{int}$ is the interaction part of the hamiltonian: 
\be
H_{int}(\tau) = \frac{g^2}{2}\, 
 \int d^dx\, \Fg^{*\,2}(\tau,\bfx) \Fg^2(\tau,\bfx),
\label{5.2}
\ee
Using the expression (\ref{3.2.2}) for $D$, this can be rewritten in the 
form
\be
\tau\, \frac{dD}{d\tau} = \frac{(2-d)}{2}\, g^2\, \dd{D}{g^2}.
\label{3.11}
\ee

\nit
{\bf 4.\ Quantum effects}
\vs{1}

\nit
The results of quantum field theory are expressed in terms of the 
time-ordered correlation functions, which can be calculated in various 
ways. In the interaction representation of the canonical formulation 
the connected parts of the correlation functions are given by
\be
G_n(z_1, ..., z_{2n}) = \cN_c \langle 0| T \lh \Fg(z_1) ...\Fg(z_n) 
 \Fg^*(z_{n+1}) ... \Fg^*(z_{2n}) e^{-i \int_{-\infty}^{\infty} d\tau\, 
 H_{int}} \rh |0 \rangle,
\label{5.1}
\ee
with $z = (\bfx,\tau)$. In the expression on the right-hand side $T$ is 
the time-ordering operator, whilst $\cN_c$ is a normalization factor 
dividing out the vacuum-to-vacuum contributions
\be
\cN_c^{-1} = 
 \langle 0| T e^{-i \int_{-\infty}^{\infty} d\tau\, H_{int}} |0 \rangle.
\label{5.2.1}
\ee 
Equivalently, the generating functional for the correlation functions 
\be 
\ba{l}
W[\eta,\eta^*] = -i \log Z[\eta,\eta^*], \\
 \\
\dsp{ \hs{2} =\, 
 \sum_n\, \int d^{d+1}z_1 ... d^{d+1}z_{2n}\, G_n(z_1,...,z_{2n})\,
 \eta^*(z_1) ...\eta^*(z_n) \eta(z_{n+1}) ...\eta(z_{2n}), }
\ea
\label{5.2.2}
\ee
can also be computed from the path integral
\be 
\ba{l} 
Z[\eta,\eta^*] = \dsp{ \int D\Fg^* D \Fg\, 
 e^{iS[\Fg,\Fg^*] + i \int_{\tau,\bfx} (\eta^* \Fg + \Fg^* \eta)}, }
\ea
\label{5.3}
\ee
with $S[\Fg,\Fg^*]$ the classical action
\be
S = \int d\tau \int d^dx \lh \frac{i}{2}\, 
 \Fg^* \stackrel{\leftrightarrow}{\der}_{\tau} \Fg 
 - \frac{1}{2}\, \bfnb \Fg^* \cdot \bfnb \Fg 
 - \frac{g^2}{2}\, |\Fg|^4 \rh.
\label{5.4}
\ee 
For consistency of presentation, in the following we take the operator 
expression (\ref{5.1}) as our starting point for the perturbative 
calculation of correlation functions as a series in powers of the coupling 
constant $g^2$. As the direct naive perturbative calculation of the 
correlation functions $G_n$ in general leads to divergences due to 
integrals over the infinite volume of momentum space, a renormalization 
procedure is necessary to obtain finite results wich can be related to 
measurable quantities. It will be argued below that in the present case 
the only modification needed in $d = 2$ is a multiplicative renormalization 
of the coupling constant 
\be 
g^2 = Z_g\, g_R^2, \hs{2} Z_g = 1 + \sum_{n\geq 1}\, z_n\, g_R^{2n},
\label{5.5}
\ee
where the coefficients $z_n$ depend on the regularization scheme and
become infinite in the limit where the regularization is removed. 
If one then calculates correlation functions order by order in the
renormalized coupling constant $g_R^2$, the resulting expressions  
remain finite upon removing the regulator. 

In principle we could regularize the theory by introducing a cut-off 
$\Lb$ in momentum space, restricting momentum integrals to the sphere
$k^2 < \Lb^2$. However, such a cut-off would introduce an explicit 
violation of the conformal symmetry (scaling and conformal boosts), 
directly spoiling the conservation of the scaling operator $D$. Instead 
in this paper we follow a different route, doing computations in $d \neq 2$ 
dimensions, treating $d$ as a continuous parameter and taking the limit 
$d \rightarrow 2$ only at the end. This procedure is closely related to 
the standard dimensional regularization used in relativistic QFT 
\ct{thooft-veltman,bollini}, except that we continue only the number 
of spatial dimensions $d$, treating the time dimension separately.  

According to eq.\ (\ref{3.5}) the scaling charge is again not conserved 
during the calculation in $d = 2 -\eps$ dimensions, be it in a very 
controled way. The question then is, whether the conservation of $D$ is 
restored in the limit $\eps \rightarrow 0$. Below we show that this is not 
the case: quantum effects spoil the scaling symmetry in $d = 2$. Indeed, 
following standard renormalization group arguments let us introduce a 
scale $\Lb$ at which the renormalized coupling $g^2_R(\Lb)$ is defined, 
such that the bare coupling remains fixed:
\be
g^2 = f(g_R^2(\Lb), \Lb, \eps), \hs{2} \frac{dg^2}{d\Lb}
 = \frac{dg_R^2}{d\Lb} \ld \dd{f}{g_R^2} \right|_{\Lb} 
 + \ld \dd{f}{\Lb} \right|_{g_R^2} = 0.
\label{5.6}
\ee
This result defines the $\bg$-function:
\be 
\bg(g_R^2, \Lb, \eps) = \Lb\, \frac{dg_R^2}{d\Lb} 
 = -\Lb\, \frac{\der f/\der\Lb}{\der f/\der g_R^2}.
\label{5.7}
\ee
Then in the renormalized theory the dimensionless scaling charge 
can be a function only of the renormalized coupling $g_R^2$ and 
the dimensionless time-parameter $y = \Lb^2 \tau$: 
\be
D = D_R(g_R^2,y).
\label{5.8}
\ee
Now as $D$ cannot depend on the renormalization scale $\Lb$, we must
have
\be
\Lb\, \frac{dD}{d\Lb} = \Lb\, \frac{dg_R^2}{d\Lb}\, \dd{D_R}{g_R^2}
 + 2 y\, \dd{D_R}{y} = 0.
\label{5.9}
\ee
It then follows that
\be
\tau\, \frac{dD}{d\tau} = y\, \dd{D_R}{y} 
 = - \frac{\bg}{2}\, \dd{D_R}{g_R^2}.
\label{5.10}
\ee
Comparing this with eq.\ (\ref{3.11}) we find that consistency requires
\be 
\frac{\eps g^2}{2}\, \dd{D}{g^2} = - \frac{\bg}{2}\, \dd{D_R}{g_R^2}.
\label{5.11}
\ee
Eqs.\ (\ref{5.10}) and (\ref{5.11}) imply, that if the $\bg$-function 
does not vanish in the limit $\eps \rightarrow 0$, then there is a 
scale anomaly proportional to the $\bg$-function; this is precisely 
what one would expect in a non-finite renormalizable quantum field theory. 
\vs{2}

\nit
{\bf 5.\ Computation of the 2- and 4-point function} 
\vs{1}

\nit
In this section we compute the quantum contributions to the propagator
and the 2-particle scattering amplitude to all orders in perturbation 
theory in $d = 2 - \eps$ dimensions. We can then compute the 
$\bg$-function, and show that equation (\ref{5.11}) becomes an 
identity. From the explicit expression for the $\bg$-function it 
follows that it does not vanish in $d = 2$, and there is a scale
anomaly. We start by defining the tree-level propagator
\be 
\ba{lll}
D_+(\tau, \bfx) & = & \dsp{ \thg(\tau)
 \int \frac{d^dk}{(2\pi)^d}\, e^{i \bfk \cdot \bfx 
 - \frac{i}{2}\, \bfk^2 \tau} }\\
 & & \\
 & = & \dsp{ \frac{i}{(2\pi)^{d+1}}\, \int_{-\infty}^{\infty} dk_0
 \int d^dk\, \frac{e^{i \bfk \cdot \bfx 
 - i k_0 \tau}}{k_0 - \frac{1}{2}\, \bfk^2 + i \ve}, }
\ea
\label{6.1}
\ee
which is a solution of the inhomogeneous free Schr\"{o}dinger
equation for positive time intervals $\tau > 0$:
\be
\lh i \der_{\tau} + \frac{1}{2}\, \Del \rh D_+(\tau,\bfx) = 
 i \del(\tau) \del^d(\bfx).
\label{6.2}
\ee
As such it represents the time-ordered two-point correlation 
function of the free theory:
\be 
\langle 0|T\lh \Fg(\tau_2,\bfx_2) \Fg^*(\tau_1,\bfx_1) \rh|0\rangle_{g^2 = 0}
 = D_+(\tau_2 - \tau_1, \bfx_2 - \bfx_1),
\label{6.3}
\ee 
which naturally vanishes for $\tau_2 < \tau_1$. In the interacting 
$|\Fg|^4$-theory the connected two-point function is defined by
\be 
G_2(z_1,z_2) = \cN_c\, \langle 0| T \lh \Fg(z_1) \Fg^*(z_2) 
  e^{-i\int_{\tau} H_{int}} \rh | 0 \rangle, 
\label{6.3.1}
\ee
where $z_i = (\tau_i, \bfx_i)$. It reduces to the tree-level result 
(\ref{6.3}) for $g^2 = 0$. The quantum corrections to the tree-level 
propagator are represented by the 1-loop diagram of fig.\ 1.

\bc
\scalebox{0.6}{\includegraphics{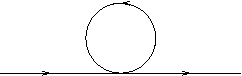}} \\
\vs{1}

{\footnotesize Fig.\ 1: 1-loop propagator correction}
\ec
\vs{1}

\nit
It is straightforward to see, that this contribution vanishes. Indeed, 
the loop represents the contraction of two field operators at the same 
point in space-time. In such an equal-time contraction, as in $H_{int}$ 
itself, the operators are normal ordered, hence their contribution to 
the vacuum-to-vacuum amplitude vanishes \ct{abrikosov}. Alternatively, 
we may evaluate the loop in the diagram
\be\ba{lll}
D_+(0, {\bf 0}) & = & \dsp{ 
 \frac{i}{(2\pi)^{d+1}}\, \int_{-\infty}^{\infty} dk_0
 \int d^dk\, \frac{1}{k_0 - \frac{1}{2}\, \bfk^2 + i \ve} }\\
 & & \\
 & = & \dsp{ \frac{1}{2 (2\pi)^d}\, \int d^dk = \frac{1}{(4\pi)^{d/2}}\,
 \frac{1}{\Gam(d/2)}\, \int_0^{\infty} dk\, \lh k^2 \rh^{\frac{d-1}{2}} = 
 0, }
\ea
\label{6.3.2}
\ee
where the last result is obtained by employing a rotation-invariant 
regularization scheme such as dimensional regularization 
\ct{thooft-veltman,bollini}. Higher-order quantum corrections consist 
either of insertions of these seagull-type propagator corrections, 
which all vanish by the argument above, or of bubbles connecting two 
propagators ($t$-channel bubbles), as in fig.\ 2:

\bc
\scalebox{0.5}{\includegraphics{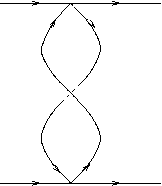}} \\
\vs{1}

{\footnotesize Fig.\ 2: $t$-channel bubble correction}
\ec
\vs{1}

\nit
Again, the time-ordering of the fields in these diagrams always forces 
these contributions to vanish; this has been made explicit in the 
figure by drawing the vertices always in a time-ordered way, with two 
incoming lines from the left, end two outgoing lines to the right. 
As a result of this analysis of propagator corrections, neither the 
fields nor chemical potential $\mu$ are renormalized by quantum loop 
effects. This shows that we can without loss of generality take 
$\mu = 0$ from the start, as we have done in (\ref{3s.1}) by switching 
to the fields $(\Fg, \Fg^*)$. 

Next we consider the connected 4-point correlation function
\be 
G_4(z_1,z_2,z_3,z_4) = \cN_c \langle 0 | T \lh \Fg(z_1) \Fg(z_2) 
 \Fg^*(z_3) \Fg^*(z_4) e^{-i\int_{\tau} H_{int}} \rh | 0 \rangle.
\label{6.4}
\ee
In view of the vanishing of the seagull-diagrams and $t$-channel bubbles, 
the full first and second order contributions in perturbation theory are 
represented by the diagrams of fig.\ 3:
\vs{1} 

\bc
\scalebox{0.6}{\includegraphics{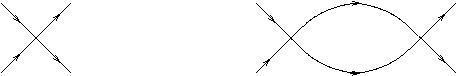}} \\
\vs{1}
{\footnotesize Fig.\ 3: 4-point diagrams} 
\ec
\vs{1}

\nit
The expression corresponding to the combined set of diagrams is
\be 
\ba{l}
\dsp{ \frac{-2ig^2}{(2\pi)^{3(d+1)}}\, \int_{k_1} ... \int_{k_4} 
 \del^{d+1}(k_1 + k_2 - k_3 - k_4)\, A^{(2)}(k_1,k_2) }\\
 \\
\dsp{ \hs{9} \times\, \prod_{i=1}^4 \lh \frac{e^{i k_i \cdot z_i}}{k_{i,0} 
 - \frac{1}{2} \bfk_i^2 + i \ve} \rh, }
\ea
\label{6.5}
\ee
where $k \cdot z = \bfk \cdot \bfx - k_0 \tau$, and 
\be 
\ba{l} 
\dsp{ A^{(2)}(k_1,k_2) = 1 - \frac{ig^2}{(2\pi)^{d+1}}\, 
 \int dk_0 \int d^d k\, B(k;k_1,k_2) }\\
 \\
B(k;k_1,k_2) = \frac{1}{\lh k_0 - \frac{1}{2} \bfk^2 + i\ve \rh 
 \lh k_0 - k_{1,0} - k_{2,0} + \frac{1}{2} \lh \bfk - \bfk_1 - \bfk_2 \rh^2
 - i \ve \rh}. 
\ea
\label{6.6}
\ee
Now we can also compute all higher-order corrections, as the only new
contribution at each order $g^{2n}$ is another loop attached to the 
diagram of order $g^{2(n-1)}$, as in fig.\ 4. 

\bc
\scalebox{0.5}{\includegraphics{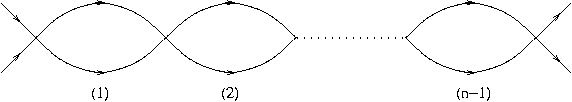}} \\
\vs{2}
{\footnotesize{Fig.\ 4: $n^{\mbox{th}}$-order contribution to $G_4$.}}
\ec 

\nit
Thus to $n^{th}$-order in the perturbative expansion of $G_4$ we get
\be 
\ba{l}
G^{(n)}_4(z_1,z_2,z_3,z_4) = \dsp{ \frac{-2ig^2}{(2\pi)^{3(d+1)}}\, 
 \int_{k_1} ... \int_{k_4} \del^{d+1}(k_1 + k_2 - k_3 - k_4)\, 
 A^{(n)}(k_1,k_2) }\\
 \\
\dsp{ \hs{9} \times\, \prod_{i=1}^4 \lh \frac{e^{i k_i \cdot z_i}}{k_{i,0} 
 - \frac{1}{2} \bfk_i^2 + i \ve} \rh, }
\ea
\label{6.6.1}
\ee
with
\be
A^{(n)}(k_1,k_2) = \sum_{p = 0}^{n-1} \lh - \frac{ig^2}{(2\pi)^{d+1}}\, 
 \int dk_0 \int d^d k\, B(k;k_1,k_2) \rh^{p}.
\label{6.6.2}
\ee
This implies that the all-order expression for the sum of bubble diagrams 
representing $G_4$ is given by a momentum amplitude $A(k_1,k_2)$ which is
the sum of a geometric series 
\be
g^2 A(k_1,k_2) \equiv g^2 A^{(\infty)}(k_1,k_2) 
 = \frac{g^2}{1 + \frac{ig^2}{(2\pi)^{d+1}}\, 
 \int dk_0 \int d^d k\, B(k;k_1,k_2)}.
\label{6.6.3}
\ee
To evaluate this expression, we first perform the $k_0$-integral by 
closing the contour in the upper-half plane; as only the pole at in 
the second factor is inside the contour, we then get after taking the 
external energies on-shell\footnote{I.e., $k_0 = \frac{1}{2} \bfk^2$.}:
\be 
A^{-1}(k_1,k_2) -1 = \frac{g^2}{(2\pi)^d}\, \int d^d k\, 
\frac{1}{\bfk^2 - \frac{1}{4} \lh \bfk_1 - \bfk_2 \rh^2 - i \ve}.
\label{6.7}
\ee
To properly perform the integral, we make the integrand and the 
integration measure dimensionless by introducing a reference momentum 
scale $\Lb$, and writing
\be
\bfk = \bfkg\, \Lb, 
\label{6.8}
\ee
where $\bfkg$ is dimensionless. Then
\be
A^{-1}(k_1,k_2) - 1 = \frac{g^2 \Lb^{d-2}}{(2\pi)^d}\, \int d^d \bfkg\, 
\frac{1}{\bfkg^2 - \frac{1}{4} \lh \bfkg_1 - \bfkg_2 \rh^2 - i \ve}.
\label{6.9}
\ee
By performing the integrals over the angles, and switching to a 
single remaining integration variable 
\be 
s = \frac{4 \bfkg^2}{(\bfkg_1 - \bfkg_2)^2},
\label{6.9.1}
\ee 
whilst continuing to $d = 2 - \eps$ dimensions, we then find that 
\be 
\ba{lll}
A^{-1}(k_1,k_2)- 1 & = & \dsp{ \frac{g^2}{4\pi} 
 \frac{\Lb^{d-2}}{\Gam(d/2)}\,  
 \lh \frac{\lh \bfkg_1 - \bfkg_2 \rh^2}{16\pi} \rh^{\frac{d-2}{2}} 
 \int_0^{\infty} ds \frac{s^{\frac{d-2}{2}}}{s - 1 - i \ve}  }\\
 & & \\
 & = & \dsp{ \frac{g^2 \Lb^{-\eps}}{4\pi} \lh \frac{2}{\eps}
  - \gam_E + i \pi - \ln \frac{(\bfk_1 - \bfk_2)^2}{16 \pi \Lb^2}  
  + {\cal O}[\eps] \rh. }
\ea
\label{6.10}
\ee
A result equivalent at one loop has been obtained in the context of the
Jackiw-Pi model in \ct{bergman-lozano}; a similar one-loop calculation for 
the fermion gas has been performed in \ct{nishida-son}. 

We can rewrite the result (\ref{6.10}) as
\be 
g^2 \Lb^{-\eps} A(k_1,k_2) = \frac{4 \pi}{\frac{4 \pi}{g^2 \Lb^{-\eps}} 
 + \frac{2}{\eps} - \gam_E + i \pi - \ln \frac{(\bfk_1 - \bfk_2)^2}{16 \pi \Lb^2}  
 + {\cal O}[\eps] }.
\label{6.10.1}
\ee
Clearly, the denominator diverges in the limit $\eps \rightarrow 0$.
This divergence can be absorbed in the coupling constant. Indeed,
noting that $g^2 \Lb^{-\eps}$ is dimensionless for any $d$,
we can introduce a dimensionless renormalized coupling constant 
$g_R^2(\Lb)$ by
\be
\frac{4\pi}{g^2 \Lb^{-\eps}} + \frac{2}{\eps} - \gam_E 
 =  \frac{4\pi}{g_R^2}.
\label{6.11}
\ee
This renormalization procedure is similar to the 
$\overline{\mbox{MS}}$-scheme in relativistic field theory. In the limit 
$\eps \rightarrow 0$ we then finally get
\be 
g^2 A(k_1,k_2) = \frac{4\pi}{\dsp{ \frac{4\pi}{g_R^2} - 
 \ln \left[ \frac{(\bfk_1 - \bfk_2)^2}{16 \pi \Lb^2} \right] + i \pi }},
\label{6.12}
\ee
which is finite for finite $g_R^2(\Lb)$ at all momenta. Observe, that the
real part vanishes for 
\be
\frac{1}{4} \lh \bfk_1 - \bfk_2 \rh^2 = 4 \pi \Lb^2\, e^{4\pi/g_R^2},
\label{6.12.1}
\ee 
whilst the imaginary part satisfies
\be
\left| g^2 A(k_1,k_2) \right|^2 = -4\,  \mbox{Im} \lh g^2 A(k_1,k_2) \rh,
\label{6.12.2}
\ee
as required by unitarity in $d = 2$. The natural scale at which to define 
the renormalized coupling constant is the particle mass: $\Lb = m$, which 
in the present conventions is normalized to unity. Defining $g^*_R 
\equiv g_R(1)$, the final result for the 4-point function then is
\be
\ba{lll}
G_4(z_1,z_2,z_3,z_4) & = & \dsp{ \frac{-2i}{(2\pi)^9}\,
\int_{k_1} ... \int_{k_4}\, \prod_{i=1}^4 \lh \frac{e^{i k_i \cdot z_i}}{
 k_{i,0} - \frac{1}{2} \bfk_i^2 + i \ve} \rh}\\
 & &  \\
 & & \dsp{ \hs{2} \times\, 
 \frac{g_R^{*\,2}\, \del^{d+1}(k_1 + k_2 - k_3 - k_4)}{\dsp 
 1 - \frac{g_R^{*\,2}}{4\pi} \ln \left[ \frac{(\bfk_1 - \bfk_2)^2}{16\pi} 
 \right] + \frac{ig_R^{*\,2}}{4}}.}
\ea
\label{6.12.3}
\ee

\nit
{\bf 6.\ Scale anomaly}
\vs{1}

\nit
Using the definition of the renormalized coupling (\ref{6.11}) we can
compute the exact $\bg$-function: 
\be 
\bg(g_R^2,\eps) = \ld \Lb\, \frac{dg_R^2}{d \Lb} \right|_{g^2,\eps} =
 \frac{g_R^4}{2\pi} - \eps g_R^2 \lh 1 + \frac{\gam_E}{4\pi}\, g_R^2 \rh\,
 \stackrel{\eps = 0}{\longrightarrow}\, \frac{g^4_R}{2\pi}.
\label{6.13}
\ee
It is easy to check, that the same result is obtained if one
only renormalizes the theory in the one-loop approximation. We 
also note that our result is equivalent to the scaling behaviour
of the 4-point coupling found in ref.\ \ct{droz-sasvari}. In particular
it implies that in dimensions $d<2$ there is a non-trivial fixed point 
\be
g_R^2 = \frac{2\pi \eps}{1 - \gam_E \eps/2}.
\label{6.13.a}
\ee 
In the following we study the limit $d = 2$ by taking 
$\eps \rightarrow 0$. Rewriting eq.\ (\ref{6.11}) in the form
\be 
g^2 =  \frac{g_R^2 \Lb^{\eps}}{\dsp{1 - \frac{g_R^2}{2\pi \eps}
 + \frac{\gam_E}{4\pi}\, g_R^2 }},
\label{6.14}
\ee
it is straightforward to show that 
\be 
\eps g^2\, \dd{}{g^2} = \lh \eps g_R^2 \lh 1 + 
 \frac{\gam_E}{4\pi} g_R^2 \rh- \frac{g_R^4}{2\pi} \rh \dd{}{g_R^2}
 = - \bg \dd{}{g_R^2},
\label{6.15}
\ee
which proves the identity (\ref{5.11}). To make the existence of 
a scale anomaly explicit, we consider the correlation function
\be 
\ba{l} 
\dsp{ \langle \Fg(z_1) \Fg(z_2) \Fg^*(z_3) \Fg^*(z_4)\, 
 \frac{dD}{d\tau} \rangle }\\
 \\
\dsp{ \hs{5} 
 \equiv \cN_c \langle 0| T \lh \Fg(z_1) \Fg(z_2) \Fg^*(z_3) \Fg^*(z_4)\,
 \frac{dD}{d\tau}\, e^{-i \int_{\tau} H_{int}} \rh |0 \rangle, }
\ea
\label{6.16}
\ee
and prove that it is non-zero. The easiest way to do this, is to 
use the result (\ref{3.10}) and integrate the equation over time:
\be
\ba{l}
\dsp{ \int d\tau\, \langle \Fg(z_1) \Fg(z_2) \Fg^*(z_3) \Fg^*(z_4)\, 
 \frac{dD}{d\tau} \rangle = \eps \int d\tau\, 
 \langle \Fg(z_1) \Fg(z_2) \Fg^*(z_3) \Fg^*(z_4)\, H_{int} \rangle }\\
 \\
\dsp{ \hs{1} = i \eps g^2 \dd{}{g^2}\, 
 \langle \Fg(z_1) \Fg(z_2) \Fg^*(z_3) \Fg^*(z_4) \rangle =
 - i \bg\, \dd{}{g_R^2}\, G_4(z_1,z_2,z_3,z_4). }
\ea
\label{6.17}
\ee
The step from the first to the second line is not quite trivial; more 
precisely, one finds
\be 
\ba{l}
\dsp{ i \eps\, g^2 \dd{}{g^2}\, \langle \Fg(z_1) \Fg(z_2) \Fg^*(z_3) \Fg^*(z_4) 
 \rangle }\\
 \\
\dsp{ \hs{3} = i \eps\, \cN_c\, g^2 \dd{}{g^2}\, \langle 0| T \lh \Fg(z_1) \Fg(z_2) 
 \Fg^*(z_3) \Fg^*(z_4) e^{-i \int_{\tau} H_{int}} \rh |0 \rangle }\\
 \\
\dsp{ \hs{4} -\, \eps\, \langle \Fg(z_1) \Fg(z_2) \Fg^*(z_3) \Fg^*(z_4) \rangle
 \int d\tau\, \langle H_{int} \rangle }\\
 \\
\dsp{ \hs{3} 
 = \eps\, \int d\tau\, \langle \Fg(z_1) \Fg(z_2) \Fg^*(z_3) \Fg^*(z_4)\,
 H_{int} \rangle, }
\ea
\label{6.18}
\ee
where we use $\langle H_{int} \rangle = 0$. Eq.\ (\ref{6.17}) shows explicitly
that the scaling charge $D$ is not conserved in the full quantum theory, the 
anomaly being proportional to the $\bg$-function. 
\vs{2}

\nit
{\bf 7.\ Discussion} 
\vs{1}

\nit
The free (linear) Schr\"{o}dinger theory in $d$ space dimensions is 
invariant under the full Schr\"{o}dinger group, and this holds for the 
non-linear classical $|\Fg|^{4}$ theory with $d = 2$ as well. As we 
have shown however, in the non-linear quantum field theory the conformal 
part of the symmetry, comprising the dilatations and special conformal 
transformations, is anomalous; the anomaly is proportional to the 
$\bg$-function:
\be 
\frac{dK}{d\tau} = - \tau \frac{dD}{d\tau} = 
 \frac{\bg}{2}\, \dd{D_R}{g_R^2}.  
\label{d.1}
\ee
Of course, the subgroup of galilean symmetries: time- and 
space-translations, rotations and galilean boosts, is still manifest
in the full theory. Other conformal field theories of the same type 
are $|\Fg|^6$ in $d = 1$, which we have not analyzed, and the 
$|\Fg|^{2}$-model for $d \rightarrow \infty$, which is a free theory 
and therefore has no anomaly. 

Taking the $\bg$-function (\ref{6.13}) and solving for the running 
coupling, again taking the mass $m$ as reference scale, we get:
\be 
\ag_R(\Lb) \equiv \frac{g_R^2(\Lb)}{2\pi} 
 = \frac{\ag_R^*}{1 - \ag_R^* \ln \Lb },
\label{d.2}
\ee
showing that the theory is infrared free and has a Landau-type
singularity (in units of $m$) for 
\be 
\Lb_s = e^{1/\ag^*_R} = \Lb\, e^{1/\ag_R(\Lb)}.
\label{d.3}
\ee
The behaviour of $\ag_R(\Lb)$ has been plotted in fig.\ 5. For 
$0 < \Lb < e^{-1} \Lb_s$ the theory is in the perturbative regime 
$0 < \ag_R < 1$. In this regime it represents an effective theory 
for a weakly interacting Bose-gas with repulsive hardcore (i.e.,  
$\del$-function) interactions at $T = 0$ in an infinite volume. It 
is important that the zero in the real part of the 4-point function 
(\ref{6.15}), or equivalently in the momentum amplitude $g^2 A(k_1,k_2)$ 
of eq.\ (\ref{6.12}), is independent of the renormalization scale; 
indeed it is reached for
\be 
\frac{1}{4} \lh \bfk_1 - \bfk_2 \rh^2 = 4 \pi \lh \Lb e^{1/\ag_R} \rh^2
 = 4 \pi\, e^{2/\ag_R^*}.
\label{d.3.1}
\ee

\bc
\scalebox{0.55}{\includegraphics{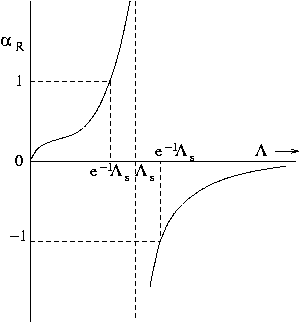}} \\
\vs{2}
{\footnotesize{Fig. 5: Running coupling of the 2-$d$ $|\Phi|^4$-model}}
\ec

\nit
It is to be noted that there is also a perturbative regime $|\ag_R| < 1$ 
for large momentum scales $\Lb > e \Lb_s$. In this regime the 
coupling constant is negative: the point-like interactions are attractive. 
Such an attractive interaction is relevant in the Jackiw-Pi model, an 
extension of the non-linear Schr\"{o}dinger model with Chern-Simons 
gauge interactions \ct{jackiw-pi,horvathy}. Indeed, the Jackiw-Pi model 
preserves the conformal symmetries of the Schr\"{o}dinger group at tree 
level, and possesses static classical solutions with zero energy in the 
regime of negative coupling. 

Observe, that replacing the bare coupling $g^2 \rightarrow -\bar{g}^2$
after renormalization in the minimal subtraction scheme leads to the
relation
\be 
\bar{g}^2\Lb^{-\eps} = \frac{\bar{g}^2_R}{\dsp{ 
 1 + \frac{\bar{g}^2_R}{2\pi \eps} 
 + \frac{\gam_E}{4\pi}\, \bar{g}_R^2}}
\label{d.4}
\ee
and a negative $\bg$-function:
\be 
\bg(\bar{g}_R^2) = - \frac{\bar{g}_R^4}{2\pi}.
\label{d.5}
\ee
The upshot is just a change of sign $\ag_R \rightarrow - \bar{\ag}_R$,
such that 
\be
\bar{\ag}_R(\Lb) = \frac{\bar{\ag}_R^*}{1 + \bar{\ag}_R^* 
 \ln \Lb}.
\label{d.6}
\ee
The behaviour of this function is as in fig.\ 5, but reflected across
the $\Lb$-axis by interchanging the positive and negative $\ag_R$-values.
This leads to exactly the same physics: repulsive interactions, now 
corresponding to $\bar{\ag}_R < 0$, in the low momentum regime below 
$\Lb_s = e^{-1/\bar{\ag}_R^*}$, and attractive interactions with 
$\bar{\ag}_R > 0$ in the high-momentum domain above $\Lb_s$. 

Of course, as a non-relativistic theory we do not expect the model to 
be valid on scales of the order $\Lb = 1$ (the mass scale), where pair 
creation becomes relevant. However, if $\Lb_s \ll 1$, there still could 
exist a perturbative regime for negative coupling $e\Lb_s < \Lb < 1$. 
Finally we observe, that in the limit $\Lb \rightarrow \infty$ the 
running coupling vanishes, as does the $\bg$-function. In this limit 
the scale invariance is restored. 

\vs{2}
\nit
{\bf Acknowledgements} \\
Discussions with K.\ Schoutens (University of Amsterdam) are gratefully acknowledged.
\vs{1}

\nit
This work is part of the research program of the Foundation for Research of Matter
(FOM), through the programs FP31 and FP52.

\np
\nit
{\bf Appendix A: the Schr\"{o}dinger group} 
\vs{1}

\nit
The Schr\"{o}dinger group is a set of space and time transformations
including space and time translations, spatial rotations, Galilei 
boosts, dilatations and special conformal transformations, acting on 
the co-ordinates $(\bfx, t)$ as 
\be
\tau^{\prime} = \frac{\ag \tau + \bg}{\gam \tau + \del}, \hs{2} 
\bfx^{\prime} = \frac{\bfR \cdot \bfx + \bfu \tau + \bfa}{\gam \tau + \del},
\label{a2.3}
\ee
where $(\ag, \bg,\gam, \del)$ are scalar parameters restricted by 
$\ag \del - \bg \gam = 1$, whilst $\bfR$ is a $d$-dimensional orthogonal 
matrix, and $(\bfu, \bfa)$ are $d$-dimensional vectors parametrizing
boosts and translations. 

The transformations (\ref{a2.3}) can be realized on the complex scalar
fields $(\Psi, \Psi^*)$ such that the free action (\ref{1.1}) remains 
invariant. This realization includes an additional $U(1)$ phase 
transformation, acting as a central charge \ct{niederer}. First, the 
subgroup of transformations with $\ag = \del = 1$ and $\gam = 0$ 
consist of time translations parametrized by $\bg$, space translations 
parametrized by $\bfa$, spatial rotations parametrized by $d(d-1)/2$ 
parameters $\og_{ab} = - \og_{ba}$ such that
\be
R_{ij} = \lh e^{ -\frac{1}{2}\, \og_{ab} L_{ab}} \rh_{ij}, \hs{2} 
(L_{ab})_{ij} = \del_{ia} \del_{jb} - \del_{ib} \del_{ja},
\label{a2.3.1}
\ee
and galilean boosts parametrized by $\bfu$. For rotations and space 
and time translations, the transformation rule for the scalar field 
is by definition
\be
\Psi(\bfx^{\prime}, \tau^{\prime}) = \Psi^{\prime}(\bfx, \tau).
\label{a2.4}
\ee
However, under galilean boosts parametrized by $\bfu$, the field 
transforms with an additional space-time dependent phase factor
\be
\Psi(\bfx^{\prime}, \tau^{\prime}) = 
 e^{i \lh \bfu \cdot \bfx + \frac{1}{2} \bfu^2 \tau \rh}\, 
 \Psi^{\prime}(\bfx, \tau).
\label{a2.5}
\ee
The remaining transformations rescale time in a non-trivial way. 
Under dilatations the field has a dimension-dependent non-zero Weyl 
weight, which also turns up in the special conformal transformations. 
Taking $\ag = 1/\del = e^\eta$, $\bg = \gam = 0$ the dilatations are 
realized on complex $\Psi$ as 
\be
\Psi(\bfx^{\prime}, \tau^{\prime}) = \ag^{-d/2} \Psi^{\prime}(\bfx, \tau)
 = e^{-d\eta/2} \Psi^{\prime}(\bfx, \tau),
\label{a2.6}
\ee
whilst the special conformal transformations with $\ag = \del = 1$
and $\bg = 0$ take the form
\be
\Psi(\bfx^{\prime}, \tau^{\prime}) = \lh 1 + \gam \tau \rh^{d/2} 
 e^{-\frac{i}{2} \frac{\gam \bfx^2}{\lh 1 + \gam \tau \rh}}\, 
 \Psi^{\prime}(\bfx,\tau).
\label{a2.7}
\ee
Finally, the $U(1)$ phase transformations leave the space-time 
co-ordinates invariant, and take the standard form on the complex
scalar field:
\be
\Psi(\bfx^{\prime}, \tau^{\prime}) = \Psi(\bfx, \tau) 
= e^{i \thg} \Psi^{\prime}(\bfx, \tau).
\label{a2.8}
\ee 
The infinitesimal forms of these transformations 
\be
\ba{lll}
\del \Psi(\bfx, \tau) & = & \Psi^{\prime}(\bfx,\tau) - \Psi(\bfx,\tau) \\
 & & \\
 & = & \bg \der_{\tau} \Psi + \bfa \cdot \bfnb \Psi + \og_{ij}\, x_i \nb_j \Psi 
 + \bfu \cdot \lh \tau \bfnb - i \bfx \rh \Psi \\
 & & \\ 
 & & \dsp{ +\, \eta \lh 2 \tau \der_{\tau} 
     + \bfx \cdot \bfnb + \frac{d}{2} \rh \Psi - i \thg \Psi }\\
 & & \\
 & & \dsp{ +\, \gam \lh - \tau^2 \der_{\tau} - \tau \lh \bfx \cdot \bfnb
     + \frac{d}{2} \rh + \frac{i}{2}\, \bfx^2 \rh \Psi, }
\ea
\label{a2.9.1}
\ee
are precisely those generated by the constants of motion 
(\ref{v.1})-(\ref{v.7}) of the free Schr\"{o}dinger theory:
\be
\ba{ll} 
\dsp{ \left\{ \Psi, H \right\} = \frac{i}{2}\, \Del \Psi 
 = \der_{\tau} \Psi, }& 
\dsp{ \left\{ \Psi, M_{ij} \right\} 
 = \lh x_i \nb_j - x_j \nb_i \rh \Psi, }\\
 & \\
\left\{ \Psi, \bfP \right\} = \bfnb \Psi, & 
\dsp{ \left\{ \Psi, D \right\} = \lh 2 \tau \der_{\tau} 
 + \bfx \cdot \bfnb + \frac{d}{2} \rh \Psi, }\\
 & \\
\left\{ \Psi, \bfG \right\} = \lh \tau \bfnb - i \bfx \rh \Psi, & 
\dsp{ \left\{ \Psi, K \right\} = \lh - \tau^2 \der_{\tau} 
 - \tau \lh \frac{d}{2} + \bfx \cdot \bfnb \rh 
 + \frac{i}{2}\, \bfx^2 \rh \Psi, }\\
& \\
\left\{ \Psi, N \right\} = -i \Psi. & 
\ea 
\label{a2.9}
\ee

\nit
{\bf Appendix B: stationary solutions} 
\vs{1}

\nit
We have proved in section 2, that in conformally invariant non-linear 
Schr\"{o}dinger models there are no non-trivial time-independent 
classical solutions. However, one can still look for stationary 
solutions, i.e.\ solutions for which the modulus of $\Psi$ is 
time-independent: 
\be
\der_{\tau} \lh \Psi^* \Psi \rh = 0.
\label{b.1}
\ee
Such fields are solutions of the time-independent non-linear 
Schr\"{o}dinger equation:
\be 
i \der_{\tau} \Psi = E \Psi, \hs{2}
- \frac{1}{2}\, \Del \Psi + \lh g^2 |\Psi|^{2(n-1)} - E \rh \Psi = 0.
\label{b.2}
\ee
In the free theory a complete set of solutions consists of the plane 
waves:
\be
\Psi_{\bfk}(\bfx,\tau) = \frac{1}{(2\pi)^{d/2}}\, 
 e^{i\bfk \cdot \bfx - i E_{\bfk} \tau}, 
\hs{2} E_{\bfk} = \frac{1}{2}\, \bfk^2. 
\label{b.0}
\ee
These solutions are $\del$-function normalized:
\be
\int d^dx\, \Psi_{\bfq}^* \Psi_{\bfk} = \del^d(\bfk - \bfq).
\label{b.0.1}
\ee
In general, also in the interacting theory there exist solutions
\be 
\Psi(\bfx,\tau) = e^{- i E \tau} \Psi(\bfx,0).
\label{b.3}
\ee
We decompose the time-independent field in modulus and phase
\be 
\Psi(\bfx,0) = a(\bfx) e^{i \thg(\bfx)}.
\label{b.4}
\ee
The second equation (\ref{b.2}) then implies two real equations
\be 
\ba{l}
2 \bfnb a \cdot \bfnb \thg + a \Del \thg = 0, \\
 \\
\dsp{ - \frac{1}{2}\, \Del a + \frac{1}{2}\, a (\bfnb \thg)^2 +
 g^2 a^{2n-1} = Ea . }
\ea
\label{b.5}
\ee
For $a \neq 0$ the first equation is equivalent to
\be 
\bfnb \cdot \lh a^2 \bfnb \thg \rh = 0.
\label{b.6}
\ee
In particular, for constant $a$ there is a set of plane-wave solutions
with
\be 
\thg(\bfx) = i \bfk \cdot \bfx, \hs{2}
E_{\bfk} = \frac{1}{2}\, \bfk^2 + g^2 a^{2(n-1)}.
\label{b.11}
\ee
To normalize these solutions we take
\be
a = \frac{1}{(2\pi)^{d/2}}, \hs{2} 
 E_{\bfk} = \frac{1}{2}\, \bfk^2 + \frac{g^2}{(2\pi)^{d(n-1)}}
 = \frac{1}{2}\, \bfk^2 + \frac{g^2}{(2\pi)^{2}},
\label{b.11.1}
\ee 
where we have used eq.\ (\ref{2.2}). Thus the plane-wave solutions
are
\be 
\Psi_{\bfk} = \frac{1}{(2\pi)^{d/2}}\, 
 e^{i \bfk \cdot \bfx - i E_{\bfk} \tau},
\label{b.12}
\ee
normalized according to the $\del$-function norm 
\be 
\int d^dx\, \Psi^*_{\bfq} \Psi_{\bfk} = \del^d(\bfk - \bfq).
\label{b.13}
\ee
This result amounts to a completeness theorem for the plane-wave 
solutions (\ref{b.12}) in the sense that any solution of the 
non-linear Schr\"{o}dinger equation can be expanded as a linear 
combination of these plane waves, even in the absence of a 
superposition principle as holds in the linear theory.

\np
\nit

\end{document}